\documentclass[aps,pra,reprint,10pt,showpacs,superscriptaddress,floatfix]{revtex4-2}
\usepackage[utf8]{inputenc}
\usepackage[T1]{fontenc}
\usepackage{amsmath,amssymb,physics}
\usepackage{bbm}
\usepackage{graphicx,booktabs}
\usepackage{orcidlink}
\usepackage{tikz}
\usepackage{pgfplots}
\pgfplotsset{compat=1.18}
\usepackage{hyperref}

\usepackage[english]{babel}

\graphicspath{{pdf/}}
\hypersetup{colorlinks=true,linkcolor=blue,citecolor=blue,filecolor=blue,urlcolor=blue,breaklinks=true}
\RequirePackage{color}

\begin{document}

\title{Geometrical optical activity induced by a continuous distribution of screw dislocations}

\author{Humberto~Belich\orcidlink{0000-0002-8795-1735}}
\email{humberto.belich@ufes.br}
\affiliation{Departamento de F\'{\i}sica e Qu\'{\i}mica, Universidade Federal do Esp\'{\i}rito Santo,\\Avenida Fernando Ferrari, 514, Goiabeiras, Vit\'oria, Esp\'{\i}rito Santo 29060-900, Brazil
}

\author{Edilberto~O~Silva\orcidlink{0000-0002-0297-5747}}
\email{edilberto.silva@ufma.br}
\affiliation{Programa de P\'{o}s-Gradua\c{c}\~{a}o em F\'{i}sica, Universidade Federal do Maranh\~{a}o, 65080-805, S\~{a}o Lu\'{i}s, Maranh\~{a}o, Brazil}

\affiliation{Coordena\c c\~ao do Curso de F\'{\i}sica -- Bacharelado, Universidade Federal do Maranh\~{a}o, 65085-580 S\~{a}o Lu\'{\i}s, Maranh\~{a}o, Brazil}
\date{\today}

\begin{abstract}
We study light propagation in a medium with uniform torsion, modeled as a continuum of screw dislocations within the geometric theory of defects. By solving Maxwell's equations in covariant form, we show that torsion induces intrinsic chirality and circular birefringence: right- and left-circular polarizations acquire different wavenumbers, leading to a purely geometric optical activity. The polarization plane of a linearly polarized beam rotates according to the simple law $\Delta\theta = \Omega\rho L$, linear in the dislocation density $\Omega$, propagation length $L$, and transverse coordinate $\rho$. This can be recast as an effective birefringence $\Delta n = 2c\Omega\rho/\omega$, providing geometric design rules for torsion-induced rotatory power. Using parameters from dislocated semiconductors, we obtain millidegree rotations over millimetre-scale paths, within reach of modern polarimetric techniques and amenable to enhancement in metamaterial platforms. We also show that the same spiral geometry implements a broadband geometric phase gate for polarization qubits and has an electronic analogue on the surface of cylindrical topological insulators, where torsion shears the Dirac cone, establishing a unified geometric link between torsion, optical activity, and topological electronic responses.
\end{abstract}

\maketitle

\section{Introduction}

The description of defects in elastic media through the methods of differential geometry has become a powerful and elegant tool, opening new perspectives in condensed matter and photonics. Inspired by the seminal ideas introduced by Kröner and Bilby \cite{Kroener1962,Bilby1955}, the geometric theory of defects establishes a direct correspondence between topological defects in solids and the geometric properties of Riemann--Cartan manifolds. In this formulation, disclinations are associated with curvature, whereas dislocations are directly related to torsion \cite{AoP.1992.216.1}. Consequently, elastic deformations induced by these defects become encoded in the spatial geometry itself, enabling the description of particle and quasiparticle dynamics as motion governed by effective curved or torsioned geometries.

A particularly interesting example is the screw dislocation, which has been extensively studied due to its impact on the electronic and optical properties of semiconductor materials and metamaterials \cite{CQG.1997.14.1129,PLA.2001.289.160}. Previous investigations have shown that the field of a single screw dislocation affects the phase of a quantum particle in a way analogous to the Aharonov--Bohm effect \cite{PR.1959.115.485}, with the dislocation's Burgers vector playing a role similar to an enclosed magnetic flux \cite{PLA.2001.289.160}. Moving beyond the scenario of isolated defects, an exact solution describing a continuous and homogeneous distribution of parallel screw dislocations was studied in Ref.~\cite{silvanetto2008}. The resulting metric, known as the spiral solution, describes a spatial geometry endowed with uniform torsion. Within this geometry, scalar particles exhibit a quantized energy spectrum known as elastic Landau levels, reinforcing the analogy between uniform torsion and a uniform magnetic field \cite{JPA.2001.34.5945,silvanetto2008,GRG.2016.48.161}.

Although several studies have extensively explored the effects of geometric torsion on scalar or spinorial particles, such as electrons within a nonrelativistic framework \cite{EPJP.2017.132.123,GRG.2016.48.161}, comparatively little attention has been given to the propagation of electromagnetic waves in media characterized by uniform torsion. It is thus natural to question how an intrinsically chiral geometry, characterized by uniform torsion, affects electromagnetic wave propagation. This question is particularly significant because, while the magnetic analogy is evident for charged particles, its manifestation for neutral fields such as light requires a detailed and systematic analysis.

In this work, we explicitly address this issue by investigating electromagnetic wave propagation within the geometry defined by the spiral solution. Our central hypothesis is that the intrinsic chirality induced by uniform torsion should manifest in observable geometric optical phenomena, such as circular birefringence or optical activity. Indeed, we demonstrate that uniform torsion introduces a nontrivial term into Maxwell's equations, leading to a direct coupling between the polarization state of light and the background torsional geometry. Specifically, we predict that a linearly polarized light beam experiences a rotation of its polarization plane upon propagating through a medium containing a uniform density of screw dislocations. Remarkably, the predicted effect displays linear dependencies on the defect density ($\Omega$), the propagation distance ($L$), and the radial distance from the symmetry axis ($\rho$), implying a differential rotation across the transverse beam profile.

Beyond its fundamental theoretical relevance, our result also suggests immediate practical applications in materials engineering. In particular, we propose that carefully designed metamaterials exhibiting effective geometric properties analogous to torsioned geometries could serve as optically active devices based solely on geometric principles.

\section{The theoretical framework \label{tf}}

We begin by considering the effective spatial geometry that models a medium with a continuous and uniform density of screw dislocations aligned along the \(z\)-axis. In cylindrical coordinates \( (x^i) = (\rho, \phi, z) \), the corresponding spatial line element is given by
\begin{equation}
    dl^2 = d\rho^2 + \rho^2 d\phi^2 + \left(dz + \Omega \rho^2 d\phi\right)^2,
    \label{eq:metric_spatial}
\end{equation}
where \( \Omega = b \sigma / 2 \) is the geometric parameter associated with the density of screw dislocations \( \sigma \), and \( b \) denotes the magnitude of the Burgers vector. This helical geometry induces an effective torsion in the medium, which directly impacts the polarization and phase velocity of propagating electromagnetic waves.

In the presence of torsion, the covariant derivative becomes modified by the contortion tensor. Specifically, the full connection is given by \( \tilde{\Gamma}^\lambda{}_{\mu\nu} = \Gamma^\lambda{}_{\mu\nu} + K^\lambda{}_{\mu\nu} \), where \( \Gamma^\lambda{}_{\mu\nu} \) is the Levi-Civita connection determined by the metric, and \( K^\lambda{}_{\mu\nu} \) is the contortion tensor, which encodes the torsional properties of the geometry. The torsion tensor \( T^\lambda{}_{\mu\nu} \) characterizes the density of dislocations and, in this geometry, the only non-vanishing component is
\begin{equation}
    T^z{}_{\phi\rho} = -T^z{}_{\rho\phi} = 2\Omega\rho.
\end{equation}
From this, the relevant component of the contortion tensor is computed using
\begin{equation}
    K^\lambda{}_{\mu\nu} = \frac{1}{2} \left( T^\lambda{}_{\mu\nu} - T_\mu{}^\lambda{}_\nu - T_\nu{}^\lambda{}_\mu \right).
\end{equation}
A straightforward calculation yields the component responsible for the coupling with the electromagnetic field:
\begin{align}
    K_{\rho\phi z} &= \frac{1}{2} \left( T_{\rho\phi z} - T_{\phi\rho z} - T_{z\rho\phi} \right) \nonumber \\
                   &= -\frac{1}{2} g_{zz} T^z{}_{\rho\phi} = \Omega\rho. 
    \label{eq:contortion_calc}
\end{align}
This term, proportional to the defect density \( \Omega \), acts as the geometric source of the chiral optical effect.

\paragraph*{\bf Derivation of the wave equation from first principles.}
To derive the wave equation for the electromagnetic potential \( A^\mu \) in a Riemann--Cartan spacetime, we begin with Maxwell’s equations in their fully covariant form:
\begin{align}
    \tilde{\nabla}_{[\lambda}F_{\mu\nu]} &= 0, \label{eq:maxwell_hom}\\[6pt]
    \tilde{\nabla}_\mu F^{\mu\nu} &= J^\nu, \label{eq:maxwell_inhom}
\end{align}
where \( \tilde{\nabla}_\mu \) denotes the covariant derivative with respect to the Cartan connection \( \tilde{\Gamma}^\lambda{}_{\mu\nu} = \Gamma^\lambda{}_{\mu\nu} + K^\lambda{}_{\mu\nu} \). In this work, we consider the source-free case, \(J^\mu = 0\), which corresponds to a region without free charges or currents. This assumption allows us to focus exclusively on the influence of the underlying torsional geometry on electromagnetic wave propagation. All observed effects, such as birefringence or optical activity, can thus be attributed solely to the background geometry rather than to conventional material responses.

The homogeneous equation \eqref{eq:maxwell_hom} ensures the existence of a four-potential \( A_\mu \), such that the field strength tensor is expressed as
\begin{equation}
    F_{\mu\nu} = \partial_\mu A_\nu - \partial_\nu A_\mu.
    \label{eq:potential_def}
\end{equation}
In this work we retain the standard, torsion-independent definition \eqref{eq:potential_def} of the field strength in terms of the four-potential. This choice keeps the usual U(1) gauge invariance manifest and implies that torsion enters Maxwell's theory only through the covariant derivatives in Eqs.~\eqref{eq:maxwell_hom} and \eqref{eq:maxwell_inhom}, rather than via additional torsion-dependent contributions to $F_{\mu\nu}$ itself.

Substituting \eqref{eq:potential_def} into the source-free equation \eqref{eq:maxwell_inhom} and adopting the generalized Lorenz gauge condition \( \tilde{\nabla}_\mu A^\mu = 0 \), we obtain the compact wave equation:
\begin{equation}
    \tilde{\nabla}_\mu\tilde{\nabla}^\mu A^\nu - \tilde{R}^\nu{}_\mu A^\mu = 0,
    \label{eq:wave_eq_compact}
\end{equation}
where \( \tilde{R}^\nu{}_\mu \) is the Ricci tensor of the Riemann--Cartan geometry.

We now expand this equation into Riemannian and torsional contributions. The d'Alembertian decomposes as
\begin{align}
    \tilde{\nabla}_\mu \tilde{\nabla}^\mu A^\nu &= \nabla_\mu \nabla^\mu A^\nu 
    + 2K^{\nu\mu\sigma} \nabla_\mu A_\sigma 
    + (\nabla_\mu K^{\nu\mu\sigma}) A_\sigma 
    \notag \\&+ (K \cdot K)^\nu{}_\sigma A^\sigma,
    \label{eq:box_expanded}
\end{align}
while the Riemann--Cartan Ricci tensor becomes
\begin{align}
    \tilde{R}^\nu{}_\mu = R^\nu{}_\mu 
    + (\nabla_\lambda K^{\nu\lambda}{}_\mu - \nabla_\mu K^{\nu\lambda}{}_\lambda) 
    + (K \cdot K)^\nu{}_\mu.
    \label{eq:ricci_expanded}
\end{align}

Inserting Eqs.~\eqref{eq:box_expanded} and \eqref{eq:ricci_expanded} into Eq.~\eqref{eq:wave_eq_compact}, we obtain:
\begin{align}
    \nabla_\mu \nabla^\mu A^\nu 
    &+ 2K^{\nu\mu\sigma} \nabla_\mu A_\sigma 
    + (\nabla_\mu K^{\nu\mu\sigma}) A_\sigma 
    - R^\nu{}_\mu A^\mu 
    \notag\\&- (\nabla_\lambda K^{\nu\lambda}{}_\mu - \nabla_\mu K^{\nu\lambda}{}_\lambda) A^\mu 
    = 0.
\end{align}

In the case of minimally coupled Maxwell theory, several torsion-dependent terms cancel identically due to symmetry properties of the field strength tensor. In particular, all terms involving \( \nabla K \) and \( K \cdot K \) cancel out, leaving the simplified exact wave equation:
\begin{equation}
    \nabla_\mu \nabla^\mu A^\nu - R^\nu{}_\mu A^\mu + 2K^{\nu\mu\sigma} \nabla_\mu A_\sigma = 0.
    \label{eq:wave_eq_simplified}
\end{equation}
This equation is the fundamental result governing the propagation of electromagnetic waves in a torsioned background. The first two terms encode the standard dynamics in curved space, while the final term introduces a direct, chirality-inducing coupling between the electromagnetic field and the background torsion.

\paragraph*{\bf Derivation of geometrical birefringence.}
We now apply the exact wave equation~\eqref{eq:wave_eq_simplified} to investigate how different light polarizations propagate in the presence of torsion. We assume a plane-wave ansatz traveling along the $z$ axis,
\begin{equation}
A_\sigma(x) = a_\sigma e^{i(k_z z - \omega t)},
\end{equation}
where $a_\sigma$ is a constant amplitude vector satisfying the transverse gauge condition $a_t = a_z = 0$. With this choice, the covariant derivative acts as a simple multiplication: $\nabla_z A_\sigma = i k_z a_\sigma$. Since the metric coefficients depend on the radial coordinate, this ansatz should be understood as a local, WKB-like description of the beam evaluated at a fixed radius $\rho$. We assume that the envelope of the field varies slowly in the transverse directions, so that radial derivatives of $A_\mu$ give subleading corrections compared with the longitudinal phase factor $e^{i k_z z}$.

In the full wave equation~\eqref{eq:wave_eq_simplified}, the Riemannian contributions $\nabla_\mu\nabla^\mu A^\nu - R^\nu{}_\mu A^\mu$ primarily yield diagonal terms in the wave-operator matrix. However, these contributions are of order $\mathcal{O}[(\Omega\rho)^3]$ or higher in the small parameter $\Omega\rho$ and are thus negligible compared with the torsional terms. The leading-order torsional term, $2K^{\nu\mu\sigma} \nabla_\mu A_\sigma$, becomes $2i k_z K^{\nu z\sigma} a_\sigma$ and governs the off-diagonal couplings.

Explicitly, evaluating the dominant components of the contortion tensor (from Eq.~\eqref{eq:contortion_calc}), we obtain the key coupling terms:
\begin{align}
\nu = \rho: && 2i k_z \Omega \rho \, a_\phi, \\
\nu = \phi: && -2i k_z \Omega \rho \, a_\rho.
\end{align}

Neglecting higher-order effects, we are left with the leading-order system of equations for the polarization amplitudes:
\begin{equation}
\begin{pmatrix}
\frac{\omega^2}{c^2} - k_z^2 & 2i k_z \Omega \rho \\
-2i k_z \Omega \rho & \frac{\omega^2}{c^2} - k_z^2
\end{pmatrix}
\begin{pmatrix} a_\rho \\ a_\phi \end{pmatrix}
=
\begin{pmatrix} 0 \\ 0 \end{pmatrix}.
\label{eq:matrix_system}
\end{equation}
Imposing the condition for nontrivial solutions, we require the determinant of the matrix to vanish. This yields the dispersion relation:
\begin{equation}
\left( \frac{\omega^2}{c^2} - k_z^2 \right)^2 = (2k_z \Omega \rho)^2. \label{dr}
\end{equation}

This is a quartic equation in $k_z$, which admits four distinct solutions:
\begin{align}
k_{z}^{(1)} &= -\Omega \rho - \sqrt{\Omega^2 \rho^2 + \omega^2 / c^2}, \label{eq:kz1} \\
k_{z}^{(2)} &= -\Omega \rho + \sqrt{\Omega^2 \rho^2 + \omega^2 / c^2}, \label{eq:kz2} \\
k_{z}^{(3)} &= +\Omega \rho - \sqrt{\Omega^2 \rho^2 + \omega^2 / c^2}, \label{eq:kz3} \\
k_{z}^{(4)} &= +\Omega \rho + \sqrt{\Omega^2 \rho^2 + \omega^2 / c^2}. \label{eq:kz4}
\end{align}

These solutions correspond to left- and right-circular polarizations propagating in forward and backward directions. The presence of torsion lifts the degeneracy among these modes, resulting in circular birefringence.

To identify the physically relevant modes, we restrict our attention to monochromatic light beams propagating predominantly in the positive $z$ direction, which requires $k_z > 0$. Under this condition:
\begin{itemize}
  \item $k_z^{(1)}$ is always negative, since it is the sum of two negative terms.
  \item $k_z^{(3)}$ is also always negative. This can be seen by rewriting it as
  \[
  k_z^{(3)} = \Omega \rho \left(1 - \sqrt{1 + \frac{\omega^2}{c^2 \Omega^2 \rho^2}}\right),
  \]
  and observing that the term in parentheses is strictly negative.
\end{itemize}
Therefore, we discard $k_z^{(1)}$ and $k_z^{(3)}$ as they correspond to backward-propagating waves.

The two forward-propagating solutions are thus:
\begin{align}
k_z^{(+)} &= +\Omega\rho + \sqrt{\Omega^2\rho^2 + \omega^2/c^2}, \\
k_z^{(-)} &= -\Omega\rho + \sqrt{\Omega^2\rho^2 + \omega^2/c^2}.
\end{align}
The difference between $k_z^{(+)}$ and $k_z^{(-)}$ represents the induced birefringence. This is a purely geometric effect arising from the background torsion.

To justify the neglect of higher-order terms, let us estimate their magnitude for realistic parameters. For instance, in gallium nitride (GaN), a widely used semiconductor, a typical screw-dislocation density is $\sigma \approx 10^9\,\text{cm}^{-2}$, and the Burgers vector is approximately $b \approx 5.18\,\text{\AA}$. These values lead to an effective parameter $\Omega = b\sigma/2 \approx 25.9\,\text{cm}^{-1} \simeq 2.6\times10^3\,\text{m}^{-1}$. For beam radii on the micrometer scale, say $\rho \sim 20\,\mu\text{m}$, we find $\Omega\rho \approx 5.2\times10^{-2}$. Consequently, the third-order term $(\Omega\rho)^3 \approx 1.4\times10^{-4}$ is negligible compared with the leading-order contribution.

These estimates confirm that retaining only the first-order terms in $\Omega$ ensures quantitative accuracy for current experimental conditions, thereby validating the robustness of our approximation scheme. Thus, the predicted optical activity emerges dominantly from the leading-order torsional coupling and constitutes a physically observable manifestation of spatial torsion.

\section{Geometrical optical activity and results \label{goa}} 

An analysis of the eigenvectors of the system \eqref{eq:matrix_system} shows that the solution $k_z^{(+)}$ corresponds to right-circularly polarized (RCP) light, while the solution $k_z^{(-)}$ corresponds to left-circularly polarized (LCP) light.

\begin{figure}[t]
\centering
\includegraphics[width=\columnwidth]{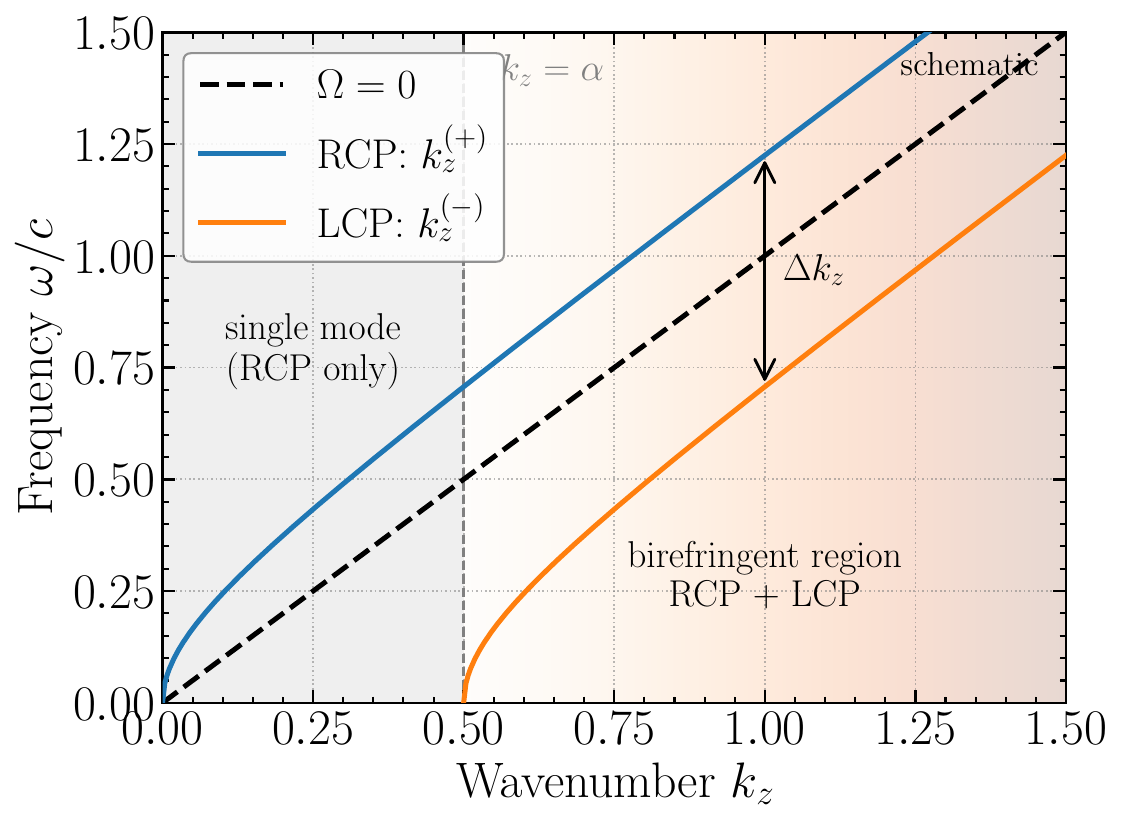}
\caption{Schematic dispersion relation $\omega/c$ versus $k_z$ in the torsioned geometry. 
The dashed black line corresponds to the torsionless case ($\Omega = 0$), where right- and left-circularly polarized modes are degenerate. In the presence of uniform torsion ($\Omega > 0$), the right-circularly polarized (RCP) and left-circularly polarized (LCP) modes follow the blue and orange branches, respectively, and acquire a nonzero splitting $\Delta k_z$. The grey region ($k_z < \alpha$) indicates the single-mode regime, where only the RCP solution is propagating, while the shaded orange region highlights the birefringent regime, where both circular polarizations propagate with different wavenumbers.}
\label{fig:birefringence_plot}
\end{figure}

The connection between circular birefringence and the rotation of linearly polarized light is a fundamental concept in optics. It begins with the understanding that a linearly polarized wave can be mathematically described as a superposition of one RCP and one LCP wave that are in phase. As the wave propagates a distance $L$, its phase evolves. Because our medium is circularly birefringent, the LCP and RCP components propagate with different wavenumbers, $k_z^{(-)}$ and $k_z^{(+)}$ respectively. Consequently, after a distance $L$, the two components will have acquired a relative phase difference of $\Delta\Phi = (k_z^{(-)} - k_z^{(+)})L$. When the two components are summed at the end of the path, they again form a linearly polarized wave, but the relative phase shift causes the plane of this new linear polarization to be rotated with respect to the original plane. The rotation angle $\Delta\theta$ is exactly half of this accumulated relative phase difference, leading to the fundamental formula
\begin{equation}
    \Delta\theta = \frac{1}{2}(k_z^{(-)} - k_z^{(+)})L.
\end{equation}
Using this formula, we can now calculate the rotation angle. The difference between the wavenumbers is
\begin{equation}
\Delta k_z = k_z^{(-)} - k_z^{(+)} = 2\Omega\rho.
\end{equation}
The rotation of the polarization plane is then:
\begin{equation}
\Delta\theta = \frac{1}{2} (\Delta k_z) L = \Omega \rho L.\label{eq:te}
\end{equation}
This result reveals three linear dependencies. The rotation is proportional to the defect density $\Omega$ and accumulates with the propagation distance $L$. Furthermore, it depends on the radial distance from the axis $\rho$, a non-trivial prediction that implies a differential rotation across a light beam.

In Fig.~\ref{fig:birefringence_plot}, we present the dispersion relation $\omega(k_z)$ illustrating the predicted geometric circular birefringence. In flat spacetime ($\Omega = 0$), the dispersion relation is linear, indicating that right- and left-circularly polarized light propagate identically. However, in the presence of uniform torsion ($\Omega > 0$), the dispersion splits into two distinct branches corresponding to the RCP and LCP polarization states. This splitting clearly demonstrates the nontrivial impact of geometric torsion, resulting in distinct propagation characteristics for different polarization states.

\begin{table*}[tbhp]
\centering
\caption{\small
Typical birefringence values ($ \Delta n $) in various optical materials at specified wavelengths. The values are obtained from experimental measurements and compiled from standard references.}
\label{tab:delta_n_materials}
\begin{tabular}{lccc}
\toprule
\textbf{Material} & \textbf{Wavelength (nm)} & $ \boldsymbol{\Delta n} $ & \textbf{Ref.} \\
\midrule
Quartz           & 589                     & $ 1.55 \times 10^{-2} $  & \cite{Hecht2002} \\
Calcite          & 589                     & $ 1.71 \times 10^{-1} $  & \cite{BornWolf1999} \\
Barium borate (BBO) & 1064                & $ 1.15 \times 10^{-1} $  & \cite{Shoji1997} \\
YVO\textsubscript{4}       & 633                     & $ 2.1 \times 10^{-1} $   & \cite{Yariv1984} \\
Lithium niobate (LiNbO\textsubscript{3}) & 633     & $ 8.3 \times 10^{-2} $   & \cite{Weis1985} \\
Mica             & 600                     & $ 3.5 \times 10^{-2} $   & \cite{BornWolf1999} \\
Liquid crystal (E7) & 633                  & $ \sim 1 \times 10^{-1} $ & \cite{Blinov2010} \\
\bottomrule
\end{tabular}
\end{table*}

\begin{figure}[t]
\centering
\includegraphics[width=\columnwidth]{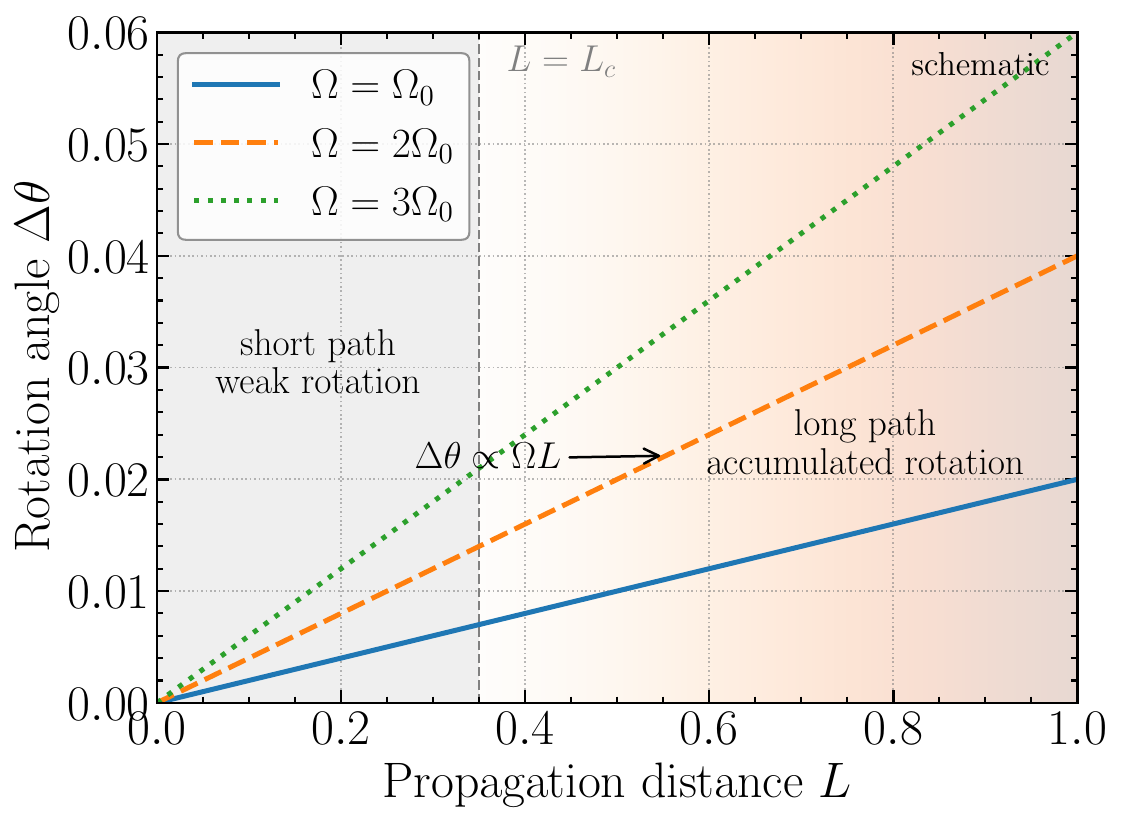}
\caption{Rotation angle $\Delta\theta$ of a linearly polarized beam as a function of the propagation distance $L$ at fixed radius $\rho$, for three values of the torsion parameter, $\Omega = \Omega_0$, $2\Omega_0$, and $3\Omega_0$. The straight lines illustrate the linear dependence $\Delta\theta \propto \Omega L$ predicted by Eq.~\eqref{eq:te}. The grey background marks the short-path regime with weak rotation, whereas the shaded orange region emphasizes the long-path regime where the rotation accumulates and becomes experimentally relevant.}
\label{fig:rotation_plot}
\end{figure}

Figure~\ref{fig:rotation_plot} further highlights the consequences of this birefringence by showing the rotation angle $\Delta\theta$ of a linearly polarized beam as a function of propagation distance $L$ for different values of torsion density $\Omega$. The linear dependence of $\Delta\theta$ on both $L$ and $\Omega$ confirms our theoretical prediction (Eq.~\eqref{eq:te}) and indicates that even subtle torsional effects accumulate into measurable polarization rotations over typical experimental propagation lengths.

While Figs.~\ref{fig:birefringence_plot} and \ref{fig:rotation_plot} already capture the essential physics, the spatially varying rotation $\Delta\theta(\rho)$ can also be visualized by a transverse cross-section of the beam, in which concentric rings at different radii carry linear polarization vectors rotated by an amount proportional to $\rho$. Such schematic representations may be particularly helpful when designing experiments aimed at imaging the radial dependence of the geometric optical activity.

\paragraph*{\bf Effective refractive indices and circular birefringence.}

The dispersion relation obtained in Eq.~\eqref{eq:matrix_system} leads directly to distinct wavenumbers $k_z^\pm$ for the right- and left-circular polarization modes:
\begin{equation}
    k_z^\pm = \pm \Omega \rho + \sqrt{\Omega^2 \rho^2 + \frac{\omega^2}{c^2}}.
\end{equation}
This allows us to define effective refractive indices for each polarization state:
\begin{equation}
    n_\pm = \frac{c\,k_z^\pm}{\omega} = \pm \frac{c \Omega \rho}{\omega} + \sqrt{\left(\frac{c \Omega \rho}{\omega}\right)^2 + 1}.
\end{equation}
These expressions describe a torsion-induced birefringence that is purely geometric in origin. The refractive index difference, which we define as
\begin{equation}
    \Delta n = n_+ - n_- = \frac{2 c \Omega \rho}{\omega},
    \label{br}
\end{equation}
is linear in the torsion strength $\Omega$ and in the beam radius $\rho$, and inversely proportional to the optical frequency $\omega$. This behavior mimics that of natural optical activity in chiral media, but here it arises solely from spacetime geometry.

\begin{table*}[ht]
\centering
\caption{\small
Predicted birefringence $ \Delta n $ using $ \Delta n = 2 c \Omega \rho / \omega $, with parameters taken from the literature: Burgers vector $ b $, dislocation density $ \sigma $, and laser wavelength $ \lambda $. The geometric torsion is calculated via $ \Omega = b \sigma / 2 $, and $ \rho = 10\,\mu\text{m} $ is fixed. All $ \lambda $ are in vacuum. References are given for the structural parameters of each material. Here $\Omega$ is quoted in $\mathrm{cm}^{-1}$ for consistency with the underlying crystallographic data.}
\label{tab:delta_n_values}
\vspace{0.3cm}
\begin{tabular}{lcccccc}
\toprule
\textbf{Material} & $ b \,(\text{nm}) $ & $ \sigma \,(\text{cm}^{-2}) $ & $ \lambda \,(\text{nm}) $ & $ \Omega \,(\text{cm}^{-1}) $ & $ \Delta n $ & \textbf{Refs.} \\
\midrule
GaN & 0.518 & $1 \times 10^9$ & 633 & 25.9 & $5.2 \times 10^{-9}$ & \cite{Ponce1999,Vurgaftman2001} \\
ZnO & 0.52 & $2 \times 10^9$ & 532 & 52.0 & $8.8 \times 10^{-9}$ & \cite{Look1998,PRB.2007.76.165202} \\
SiC & 0.25 & $5 \times 10^8$ & 532 & 6.25 & $1.1 \times 10^{-9}$ & \cite{HunterNeudeck2002} \\
InN & 0.57 & $1 \times 10^9$ & 780 & 28.5 & $7.1 \times 10^{-9}$ & \cite{SM.2003.34.63} \\
AlN & 0.498 & $8 \times 10^8$ & 633 & 19.92 & $4.0 \times 10^{-9}$ & \cite{Iqbal2018} \\
\bottomrule
\end{tabular}
\end{table*}

\paragraph*{\bf Typical birefringence values in optical media.} To contextualize our theoretical result within experimental reality, we present in Table~\ref{tab:delta_n_materials} a compilation of birefringence values ($ \Delta n = n_+ - n_- $) for several commonly studied materials. These values are taken from experimental studies on optical anisotropy in uniaxial and biaxial crystals.

Our theoretical model predicts birefringence of the form
\begin{equation}
    \Delta n = n_+ - n_- = 2 \frac{c \Omega \rho}{\omega},    
\end{equation}
which, for realistic torsion values $ \Omega \sim 10^3\text{--}10^4 \,\mathrm{m}^{-1} $, optical frequencies $ \omega \sim 10^{15} \,\mathrm{Hz} $, and typical beam widths $ \rho \sim 10^{-6}\text{--}10^{-5} \,\mathrm{m} $, yields birefringence values of order $ 10^{-9}\text{--}10^{-8} $ for the parameter ranges considered in Table~\ref{tab:delta_n_values}. Although much smaller than the birefringence of strongly anisotropic crystals, these values are in principle accessible to modern high-sensitivity polarimetric techniques.

\paragraph*{\bf Example: gallium nitride (GaN).}
For GaN crystals, which exhibit high screw-dislocation densities ($ \sigma \sim 10^9 \, \mathrm{cm}^{-2} $) and a Burgers vector $ b \sim 5.2\,\mathrm{\AA} $, the resulting torsion parameter is estimated as $ \Omega \sim 2.6 \times 10^3 \,\mathrm{m}^{-1} $. Substituting this into our formula with $ \rho = 20\,\mu\mathrm{m} $ and $ \lambda = 633\,\mathrm{nm} $, we obtain a birefringence $ \Delta n \sim 1.0 \times 10^{-8} $. The corresponding rotation angle from Eq.~\eqref{eq:te} is
\begin{align}
    \Delta\theta &\approx 3.0\times10^{-3}~\mathrm{deg}
    \left(\frac{L}{1~\mathrm{mm}}\right)
    \left(\frac{\rho}{20~\mu\mathrm{m}}\right)\notag\\&\times
    \left(\frac{\Omega}{2.6\times10^3~\mathrm{m}^{-1}}\right),
    \label{eq:GaN_rotation_estimate}
\end{align}
which, although much smaller than the rotatory powers of strongly birefringent crystals listed in Table~\ref{tab:delta_n_materials}, still lies within the reach of state-of-the-art polarimetric measurements, especially in samples with enhanced effective torsion or increased propagation length.

These estimates demonstrate that the geometric birefringence predicted by our model, while numerically small, lies within experimentally accessible regimes and can be amplified in suitably designed systems.

\paragraph*{\bf Predicting birefringence in real materials from the geometric model.} To illustrate the predictive capability of our geometric model, we apply the analytical formula~\eqref{br} for birefringence where $\rho$ is the transverse distance from the beam axis, and $\omega = 2\pi c / \lambda$ is the optical angular frequency.

Table~\ref{tab:delta_n_values} shows birefringence values calculated for a selection of technologically relevant materials using our geometric model. The material-specific parameters, including Burgers vector $b$, dislocation density $ \sigma $, and wavelength $\lambda$, are obtained from experimental reports in the literature. For consistency, we assume a representative beam radius $\rho = 10\,\mu\text{m}$. The results demonstrate that the predicted birefringence is in the range of $10^{-9}$ to $10^{-8}$. Although this range is several orders of magnitude smaller than the birefringence found in strongly anisotropic crystals, it should still be detectable with modern, high-sensitivity polarimetric setups, particularly in systems where $\Omega$ or the effective path length can be enhanced.

These results confirm that torsion-induced birefringence in realistic systems, though subtle, can be brought into the realm of experimental observation. Our model, based on the effective geometric torsion from lattice defects, thus provides a predictive and physically meaningful framework applicable to real materials.

\paragraph*{\bf Application to chiral media and metamaterials.}

The central result of this work, expressed in Eq.~\eqref{eq:te}, establishes a bridge between a microstructural property of a medium, the screw-dislocation density $\Omega$, and a macroscopic optical response---optical activity. This connection allows us to apply our geometric model to describe a broader class of materials, such as general chiral media and metamaterials.

Many studies on metamaterials follow a path of designing a specific structure and then simulating or measuring its chiral response. This response is often characterized by a phenomenological chirality parameter. Our model takes the inverse path: we start from a fundamental geometric principle (a space with uniform torsion, $\Omega$) and derive the optical response ($\Delta\theta = \Omega \rho L$). This provides a first-principles geometric origin for the phenomenological behavior of chiral media.

\paragraph*{\bf A new design principle.}
This approach offers a new design principle for materials with tailored optical properties. The most promising application lies in the field of metamaterials. While controlling the defect density in a crystal is challenging, it is entirely feasible to fabricate artificial structures at the micro- or nanoscale whose effective geometry, from the perspective of an electromagnetic wave, mimics that of our helical spacetime. For an engineer or materials scientist seeking to create a device with a specific optical activity, our model suggests that they could design a structure whose effective geometry behaves like a space with a specific torsion $\Omega$. The formula $\Delta\theta = \Omega \rho L$ would then serve as a predictive design equation to engineer the metamaterial's behavior, paving the way for novel optical devices based on purely geometric principles.

\paragraph*{\bf Extension to topological semimetals.}

Beyond engineered metamaterials, our model also applies naturally to topological semimetals, such as Dirac and Weyl semimetals, in which the low-energy electronic excitations obey relativistic-like wave equations. In these systems, lattice dislocations and strain fields can induce emergent geometrical backgrounds, including effective torsion fields~\cite{cortijo2015dislocations,pikulin2016chiral}.

A continuous distribution of screw dislocations in a semimetal creates an effective torsion similar to the one modeled by our helical geometry. This implies that the torsion-induced birefringence and polarization rotation predicted here are not merely theoretical constructs, but may manifest in real materials with high defect densities.

For example, Weyl semimetals like TaAs and NbAs typically exhibit screw-dislocation densities on the order of $10^9$ to $10^{10}\,\text{cm}^{-2}$. Combined with Burgers vectors of $b \sim 0.5\,\text{nm}$, this yields torsion parameters $\Omega \sim 25$--$100\,\text{cm}^{-1}$, similar to those estimated for GaN. In such materials, the rotation angle $\Delta\theta$ could be measurable for beam radii of a few micrometers and sample thicknesses of $L \sim 100\,\mu\text{m}$.

These results suggest that geometric optical activity can serve as a novel probe of dislocation networks in topological semimetals. Conversely, semimetals may provide a natural condensed-matter platform for observing torsion-induced optical phenomena, complementing engineered metamaterials with topological protection.

\section{Geometric design rules for torsion-induced optical activity}
\label{sec:design_rules}

The analytical expressions obtained in Sec.~\ref{tf} for the torsion-induced
birefringence and polarization rotation,
Eqs.~\eqref{br} and~\eqref{eq:te}, can be recast in the standard language of
optical activity. In a conventional optically active medium, the rotation angle
of the polarization plane after propagation over a distance $L$ is written as
\begin{equation}
    \Delta\theta
    = \frac{\pi L}{\lambda}\,\Delta n,
    \label{eq:standard_rotation}
\end{equation}
where $\lambda = 2\pi c/\omega$ is the vacuum wavelength and
$\Delta n = n_+ - n_-$ is the refractive-index difference between right-
and left-circularly polarized modes. Combining Eq.~\eqref{eq:standard_rotation}
with our geometric result~\eqref{br},
\begin{equation}
    \Delta n = n_- - n_+ = 2\frac{c\Omega\rho}{\omega},
\end{equation}
we immediately recover the purely geometric rotation law
$\Delta\theta = \Omega\rho L$ from Eq.~\eqref{eq:te}. This shows that the spiral
geometry implements the standard optical-activity formula with an effective,
geometry-induced $\Delta n$ which is fully determined by the Burgers-vector
density and the transverse position.

\subsection{Geometric rotatory power and effective chirality parameter}

Equation~\eqref{eq:te} suggests defining a \emph{geometric rotatory power}
(or torsional Verdet-like constant) as
\begin{equation}
    \mathcal{V}_{\rm geom}(\rho)
    = \frac{\Delta\theta}{L}
    = \Omega\rho,
    \label{eq:Vgeom_def}
\end{equation}
which depends linearly on the Burgers-vector density and on the transverse
coordinate. In effective-medium terms, it is also convenient to introduce a
geometric chirality parameter $\kappa_{\rm geom}$ by comparing the birefringence
of bi-isotropic chiral media, $\Delta n = 2\kappa$,
with Eq.~\eqref{br}. This yields
\begin{equation}
    \kappa_{\rm geom} = \frac{c\Omega\rho}{\omega},
    \label{eq:kappa_geom}
\end{equation}
providing a direct dictionary between the Riemann--Cartan geometry and
phenomenological models of chiral metamaterials and bi-isotropic media
used in polarization-control applications
\cite{SPRodrigues2022_review}.

From a design perspective, Eqs.~\eqref{eq:Vgeom_def} and~\eqref{eq:kappa_geom}
can be read from right to left: a target rotation per unit length
$\mathcal{V}_{\rm geom}^{\ast}$ or target chirality $\kappa_{\rm geom}^{\ast}$
immediately fixes the required value of $\Omega\rho$,
\begin{equation}
    \Omega\rho = \mathcal{V}_{\rm geom}^{\ast}
    = \frac{\omega}{c}\,\kappa_{\rm geom}^{\ast},
\end{equation}
which, through $\Omega = b\sigma/2$, translates into a constraint on the
combination $b\sigma$ characterizing the density of screw-like microstructures.
In engineered platforms this constraint can be implemented either by changing
the actual dislocation density in a crystal or, more realistically, by tailoring
the unit cell of a metamaterial so that its effective optical response mimics
the spiral geometry with a prescribed $\Omega$.

It is useful to distinguish three regimes:
(i) a \emph{perturbative torsion} regime, $|\Omega\rho|\ll \omega/c$, in which
$\Delta n \propto \Omega\rho/\omega$ is small but highly controllable and can
be treated as a linear correction to the host material;
(ii) a \emph{competitive} regime in which $|\Delta n_{\rm geom}|$ becomes
comparable to the intrinsic birefringence of the background medium, enabling
geometry to either enhance or partially cancel the natural optical activity;
and (iii) an \emph{engineered} regime where an effectively large $\Omega$
is realized in a metamaterial, pushing the geometric contribution to the
dominant role and allowing the optical response to be governed primarily by
torsion rather than by conventional material anisotropy.

\subsection{Realistic estimates for photonic and crystal platforms}

The geometric theory is particularly relevant in view of recent experiments
where torsion stresses or screw dislocations have been shown to induce optical
activity and optical anisotropy in bulk crystals and nanostructures.
Torsion-induced rotation has been reported, for instance, in centrosymmetric
NaBi(MoO$_4$)$_2$ crystals \cite{Vasylkiv2013_NaBi}, where the optical gyration
is proportional to the torsion-induced stress gradient, and in LiNbO$_3$
crystals, where twisting gives rise to optical vortices and spatially varying
birefringence \cite{Skab2011_LiNbO3}. At the nanoscale, screw dislocations in
semiconducting nanocrystals have been identified as a major source of giant
optical activity and circular dichroism
\cite{Baimuratov2015_SciRep,Ciobanu2023_MatHoriz}.

To connect with these systems, we use the estimates summarized in
Table~\ref{tab:delta_n_values}, where $\Omega = b\sigma/2$ is calculated from
reported Burgers vectors and dislocation densities. For definiteness, let us
consider a GaN crystal with $\sigma \simeq 10^9~\mathrm{cm}^{-2}$ and
$b \simeq 0.52~\mathrm{nm}$, yielding
$\Omega \simeq 2.6\times10^3~\mathrm{m}^{-1}$, and take a probe beam with
$\lambda = 633~\mathrm{nm}$ and radius $\rho = 20~\mu\mathrm{m}$
\cite{Ponce1999,Vurgaftman2001}. From Eq.~(\ref{br}), we obtain a geometric birefringence of order
$\Delta n \sim 10^{-8}$ for the GaN parameters quoted above, and
Eq.~(26) then yields the rotation angle
\begin{align}
    \Delta\theta &\approx 3.0\times 10^{-3}~\mathrm{deg}\,
    \left(\frac{L}{1~\mathrm{mm}}\right)
    \left(\frac{\rho}{20~\mu\mathrm{m}}\right)\notag\\& \times
    \left(\frac{\Omega}{2.6\times 10^{3}~\mathrm{m}^{-1}}\right).
    \label{eq37}
\end{align}
Although this corresponds to a rotation of only a few
millidegrees over millimeter-scale paths, such angles lie well
within the sensitivity of modern high-precision polarimetric
setups, especially when combined with longer propagation lengths,
multi-pass geometries, or resonant enhancement in cavities.
In this sense, the spiral geometry provides a concrete example
where a purely geometric torsion field produces a weak but
experimentally accessible optical activity in realistic
semiconductor platforms.

\begin{figure}[t]
\centering
\includegraphics[width=\columnwidth]{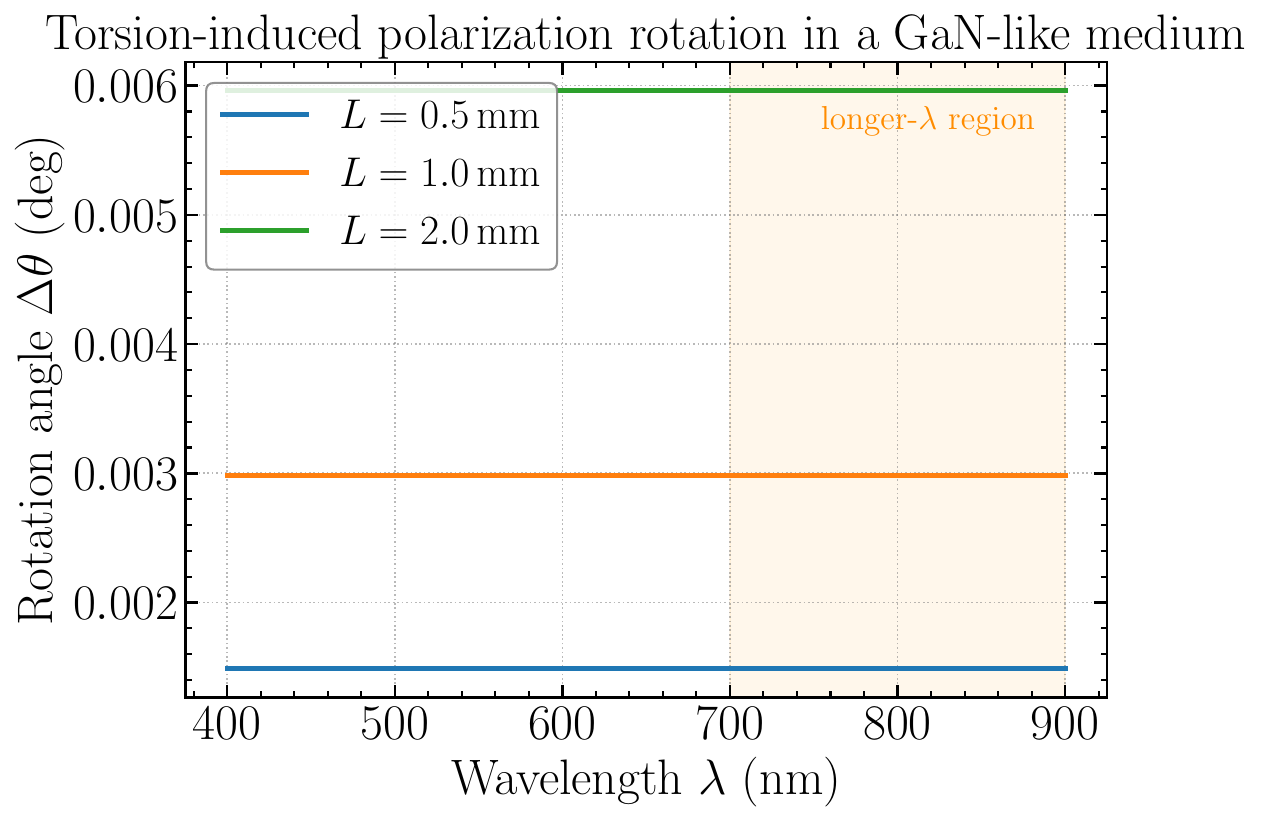}
\caption{Torsion-induced rotation angle $\Delta\theta$ as a function of
wavelength $\lambda$ for a GaN-like medium with screw-dislocation density
$\sigma = 10^9~\mathrm{cm}^{-2}$ and beam radius
$\rho = 20~\mu\mathrm{m}$, for different sample thicknesses $L$.
The curves are obtained from Eqs.~\eqref{br} and~\eqref{eq:standard_rotation},
using material parameters extracted from Refs.~\cite{Ponce1999,Vurgaftman2001}.}
\label{fig:GaN_rotation_lambda}
\end{figure}

Figure~\ref{fig:GaN_rotation_lambda} illustrates the behaviour of the rotation angle $\Delta\theta$ as a
function of the wavelength $\lambda$, obtained by combining
Eq.~(\ref{eq:standard_rotation}) with the geometric birefringence
in Eq.~(\ref{br}) for representative dislocation densities. For fixed $L$ and
$\rho$, the curves are essentially flat, reflecting the fact that in our model
the torsion-induced optical activity is \emph{achromatic}:
the explicit $\lambda$-dependence in $\Delta n(\lambda)$ exactly cancels the
$\lambda$-dependence in Eq.~(\ref{eq:standard_rotation}), yielding the purely
geometric law $\Delta\theta = \Omega \rho L$ [see Eq.~(\ref{eq:GaN_rotation_estimate})].
In the GaN-like example shown, the rotation angles lie in the range of a few
$10^{-3}$~deg over sub-millimetre and millimetre path lengths, in agreement with
Eq.~(\ref{eq:GaN_rotation_estimate}). Although small, these achromatic rotations
are within reach of state-of-the-art polarimetric techniques and can be
enhanced either by increasing the effective torsion (for instance, through
engineered dislocation networks or metamaterial unit cells) or by using
multi-pass and cavity-enhanced configurations.

In this context, the continuum torsion parameter $\Omega$ should
be understood as an effective description of a structured waveguide
or metasurface whose unit cell emulates the spiral geometry.
The design rule $\Delta\theta = \Omega\rho L$ then plays the same
role as phenomenological rotatory-power formulas in chiral
metamaterials, but with the advantage that the microscopic origin
of the optical activity is explicitly geometric and directly linked
to the Burgers-vector density and transverse position.

\section{Polarization qubits and chiral quantum optics}
\label{sec:qubits_chiral}

The circular birefringence induced by uniform torsion also admits a natural
interpretation in the language of quantum optics, where polarization encodes a
photonic qubit. Let $\ket{R}$ and $\ket{L}$ denote the right- and left-circular
polarization states. At fixed radius $\rho$, a monochromatic beam propagating
along the $z$ direction in the spiral geometry acquires different propagation
constants $k_z^{(+)}$ and $k_z^{(-)}$ for the two circular components, which
results in a relative phase
\begin{equation}
    \Delta\Phi(L)
    = \bigl(k_z^{(-)} - k_z^{(+)}\bigr)L
    = 2\Omega\rho L,
\end{equation}
after crossing a slab of thickness $L$. The evolution of the polarization
qubit is thus described by the unitary operator
\begin{equation}
    U(\Omega,\rho,L)
    = \exp\!\left[-\frac{i}{2}\Delta\Phi(L)\,\sigma_3\right]
    = \exp\!\left[-i\,\Omega\rho L\,\sigma_3\right],
    \label{eq:U_geom_gate}
\end{equation}
where $\sigma_3$ is the Pauli matrix in the $\{\ket{R},\ket{L}\}$ basis.
Equation~\eqref{eq:U_geom_gate} represents a pure \emph{geometric phase gate}:
it implements a rotation around the $z$ axis of the Poincar\'e sphere by an
angle $2\Omega\rho L$.

Because $\Omega$ is frequency independent and $\rho$ and $L$ are purely
geometric parameters, the gate in Eq.~\eqref{eq:U_geom_gate} is inherently
broadband: once a given device is fabricated with a target value of
$\Omega\rho L$, the induced phase shift is essentially dispersionless over
the transparency window of the host material. In particular, choosing
$\Omega\rho L = \pi/4$ yields a $\pi/2$ phase-shift gate,
$U = \mathrm{diag}(e^{-i\pi/4},e^{+i\pi/4})$, while
$\Omega\rho L = \pi/2$ produces a Pauli-$Z$ gate that flips the relative phase
between $\ket{R}$ and $\ket{L}$. In combination with linear-optical elements
that implement $X$ and $Y$ rotations, such torsion-based $Z$ rotations enable
universal control of polarization qubits.

For realistic parameters, the required product $\Omega L$ is compatible with
integrated-photonic footprints. Taking, for example,
$\Omega \sim 10^3$--$10^4~\mathrm{m}^{-1}$ and
$\rho \sim 5$--$10~\mu\mathrm{m}$, a $\pi/2$ gate requires effective path
lengths in the sub-millimeter to millimeter range, which can be realized in
compact waveguide spirals or ring resonators. In such architectures the
torsion-mimicking region behaves as a passive polarization phase shifter
whose action is fixed by geometry and does not rely on electro-optic or
thermo-optic tuning.

\subsection{Gate fidelity and robustness against parameter fluctuations}

\begin{figure}[t]
\centering
\includegraphics[width=\columnwidth]{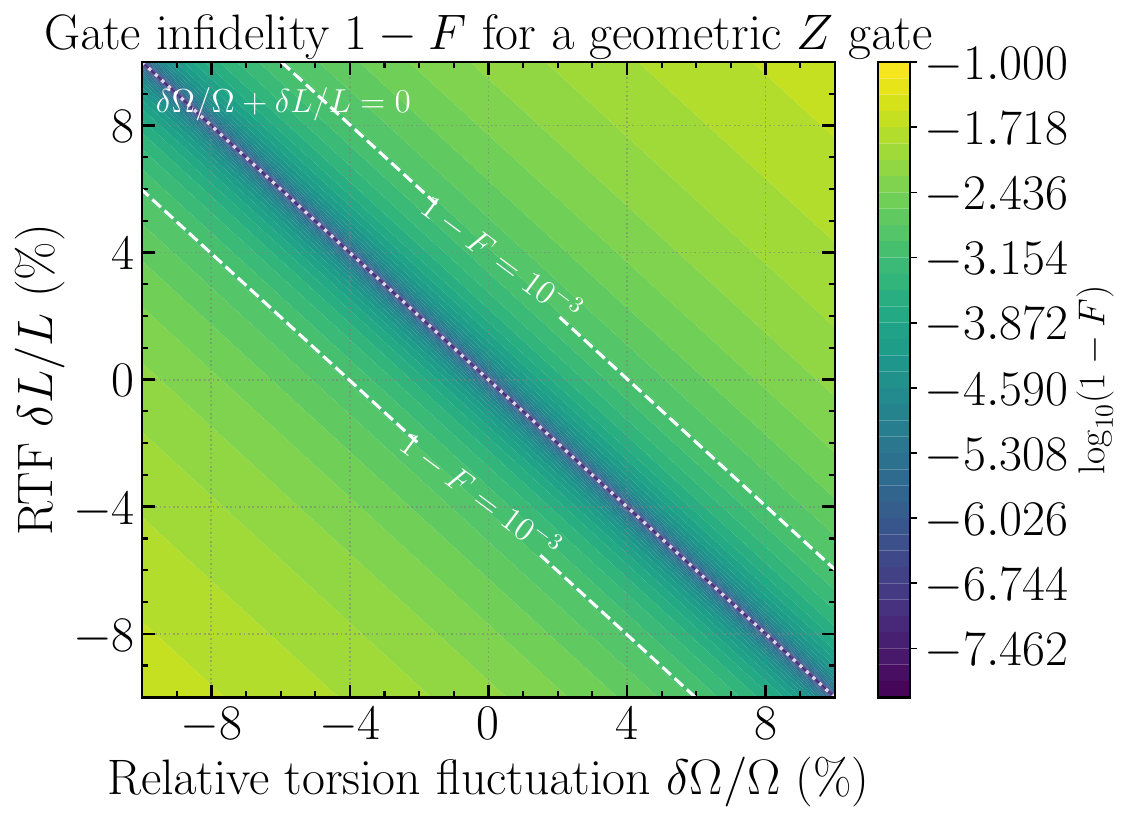}
\caption{Infidelity $1-F$ of the geometric phase gate
$U(\Omega,\rho,L)$ [RTF - Relative thickness fluctuation] as a function of relative fluctuations in the torsion
strength $\delta\Omega/\Omega$ and sample thickness $\delta L/L$, for a
target $Z$ gate ($\theta = \Omega\rho L = \pi/2$). The dashed contour marks
the $1-F = 10^{-3}$ line.}
\label{fig:fidelity}
\end{figure}

From a quantum-information perspective, it is important to assess the
sensitivity of the geometric phase gate~\eqref{eq:U_geom_gate} to fluctuations
in the parameters $\Omega$, $\rho$, and $L$. Writing
$\Omega\rho L = \theta + \delta\theta$, where $\theta$ is the target rotation
angle and $\delta\theta$ a small deviation, the corresponding unitary can be
expanded as
\begin{equation}
    U(\theta + \delta\theta)
    \simeq U(\theta)\left[\mathrm{1}
    - i\,\delta\theta\,\sigma_3
    - \frac{(\delta\theta)^2}{2}\,\mathrm{1}
    + \mathcal{O}(\delta\theta^3)\right].
\end{equation}
For an initial state $\ket{\psi}$, the average gate fidelity with respect to
the ideal operation $U(\theta)$ is then
\begin{equation}
    F \simeq 1 - \frac{(\delta\theta)^2}{4}
    = 1 - \frac{1}{4}\bigl(\delta\Omega\,\rho L 
      + \Omega\,\delta\rho\,L
      + \Omega\rho\,\delta L\bigr)^2.
    \label{eq:fidelity}
\end{equation}
Figure~\ref{fig:fidelity} shows $1-F$ as a function of relative fluctuations
$\delta\Omega/\Omega$ and $\delta L/L$ for a target $Z$ gate
($\theta = \pi/2$). The contour plot reveals that gate infidelities below $10^{-3}$ are achievable with modest control accuracies at the percent level, indicating that torsion-based phase gates can be robust against realistic fabrication and alignment tolerances in integrated photonics.

Importantly, the dependence on $\Omega\rho L$ implies that different sources of imperfections can be traded against each other: tighter control of the waveguide cross-section (and thus of $\rho$) can compensate for larger uncertainties in the effective torsion, and vice versa. This flexibility is particularly attractive in metamaterial or dislocation-engineered platforms, where $\Omega$ may be tunable only within a certain range, but the optical mode profile can be tailored with high precision.

\subsection{Connections to chiral quantum optics and topological photonics}

Recent advances in chiral and topological photonics have shown that spin-momentum locking in nanophotonic waveguides and metasurfaces can be used to route single photons in a spin-dependent way and to engineer nonreciprocal light--matter interfaces and robust photonic states \cite{Ozawa2019_RMP}. In those platforms, the polarization degree of freedom is coupled to the propagation direction through transverse spin and topological band-structure properties.

Although our model describes a homogeneous bulk medium rather than a specific waveguide geometry, the effective SU(2) rotation~\eqref{eq:U_geom_gate} can be interpreted as a building block for chiral quantum-optical devices. A segment of torsion-mimicking material inserted into a spin--momentum-locked waveguide
would implement a polarization-dependent phase shift between counterpropagating modes, thereby enabling directional phase control and nonreciprocal responses.
In this setting, the geometric parameter $\Omega$ plays a role analogous to that of an effective spin--orbit coupling strength, but with a purely geometrical origin.

Furthermore, combining geometric phase gates with localized emitters in topological photonic structures \cite{Ozawa2019_RMP} opens the possibility of spin-dependent quantum networks in which the effective torsion acts as a tunable knob for polarization and path control. For example, embedding a torsion-engineered region inside a topological waveguide or ring resonator could realize a robust, disorder-insensitive polarization gate whose action is protected by both geometry and band topology.

In this broader context, the spiral geometry offers not only a conceptually transparent realization of circular birefringence but also a geometric platform for implementing polarization-phase gates and for interfacing photonic qubits with structured media exhibiting effective torsion, dislocation networks, or topological band structures.

\section{Electronic analogue: surface Dirac fermions and topological insulators}
\label{sec:TI_connection}

Topological insulators (TIs) provide a paradigmatic platform where band topology and real-space geometry intertwine. Three-dimensional strong TIs host gapless helical Dirac fermions on their surfaces, described at low energies by a relativistic Dirac Hamiltonian with Fermi velocity $v_{\rm F}$ and spin--momentum locking \cite{HasanKane2010_RMP,QiZhang2011_RMP}. It has been shown that crystalline defects such as dislocations act as sensitive probes of the bulk topology: dislocation lines in strong TIs can bind one-dimensional topologically protected modes, whose existence is fixed by the band-structure invariant and the Burgers vector of the defect \cite{TeoKane2010_PRB,RanVishwanath2009_NatPhys,Juricic2012_PRL,Slager2014_PRB}. In parallel, work on Weyl and Dirac semimetals has demonstrated that strain and dislocations can generate emergent gauge and torsion fields, with notable consequences such as torsion-induced chiral transport \cite{SumiyoshiFujimoto2016_PRL}.

These developments suggest that the same spiral geometry used in this work to describe a medium with a continuous density of screw dislocations should admit a natural electronic analogue: a TI whose surface states propagate on a torsionful background. In this section, we sketch such an electronic realization by coupling a Dirac surface cone to the spiral metric~\eqref{eq:metric_spatial}. The result is a simple, fully analytical model in which uniform torsion mixes longitudinal and azimuthal momenta of the surface Dirac fermions, in close analogy with the purely geometric optical activity derived in the previous sections.

\subsection{Dirac surface states on a spiral cylinder}

We consider a strong three-dimensional TI whose macroscopic crystal shape is a cylinder of radius $R$ aligned with the $z$ axis. In the absence of lattice defects, the low-energy electronic states on the cylindrical surface are described by a two-dimensional Dirac Hamiltonian,
\begin{equation}
    H_0
    = \hbar v_{\rm F}
      \bigl( \sigma_\phi\,k_\phi + \sigma_z\,k_z \bigr),
    \label{eq:H0_flat}
\end{equation}
where $v_{\rm F}$ is the Fermi velocity, $(k_\phi,k_z)$ are the wave-vector components along the azimuthal and longitudinal directions, and $\sigma_\phi,\sigma_z$ are Pauli matrices acting on the spin (or effective pseudospin) degree of freedom \cite{HasanKane2010_RMP,QiZhang2011_RMP}. For a cylindrical surface of radius $R$, the single-valuedness of the wave function enforces
\begin{equation}
    k_\phi = \frac{m}{R}, 
    \qquad m \in \mathbb{Z},
\end{equation}
up to possible Berry-phase and flux corrections that we neglect here for simplicity \cite{TeoKane2010_PRB}.

We now embed this cylindrical surface in the spiral geometry describing a continuous and homogeneous distribution of screw dislocations. The three-dimensional spatial line element is
\begin{equation}
    dl^2 
    = d\rho^2 + \rho^2 d\phi^2 + \bigl(dz + \Omega \rho^2 d\phi\bigr)^2,
    \label{eq:spiral_metric_TI}
\end{equation}
identical to Eq.~\eqref{eq:metric_spatial}, with $\Omega = b\sigma/2$ determined by the Burgers vector $b$ and the areal density $\sigma$ of screw dislocations. Restricting to the cylindrical surface at fixed radius $\rho = R$, we obtain the induced spatial metric in the $(\phi,z)$ coordinates,
\begin{align}
    dl_{\rm surf}^2
    &= R^2 d\phi^2 + \bigl(dz + \Omega R^2 d\phi\bigr)^2
    \notag\\[4pt]
    &= dz^2 
       + \bigl( R^2 + \Omega^2 R^4 \bigr) d\phi^2
       + 2 \Omega R^2\,dz\,d\phi.
    \label{eq:induced_metric_surface}
\end{align}
A convenient orthonormal coframe for this surface is provided by the one-forms
\begin{equation}
    e^1 = R\,d\phi,
    \qquad
    e^2 = dz + \Omega R^2 d\phi,
    \label{eq:coframe_surface}
\end{equation}
which satisfy $dl_{\rm surf}^2 = (e^1)^2 + (e^2)^2$. The associated local orthonormal momenta $(p_1,p_2)$ are defined by expanding the phase of a Bloch state $\psi \sim e^{i(m\phi + k_z z)}$ as
\begin{equation}
    dS = m\,d\phi + k_z\,dz = p_1 e^1 + p_2 e^2.
    \label{eq:phase_expansion}
\end{equation}
Substituting Eq.~\eqref{eq:coframe_surface} into Eq.~\eqref{eq:phase_expansion}, we find
\begin{align}
    m\,d\phi + k_z\,dz
    &= p_1 R\,d\phi + p_2\bigl(dz + \Omega R^2 d\phi\bigr)
    \notag\\[4pt]
    &= \bigl(p_1 R + p_2 \Omega R^2\bigr) d\phi + p_2\,dz.
\end{align}
Matching coefficients of $d\phi$ and $dz$ gives
\begin{equation}
    p_2 = \hbar k_z,
    \qquad
    p_1 R + p_2 \Omega R^2 = \hbar m,
\end{equation}
so that
\begin{equation}
    p_1
    = \hbar\left( \frac{m}{R} - \Omega R k_z \right),
    \qquad
    p_2 = \hbar k_z.
    \label{eq:local_momenta}
\end{equation}

In the absence of additional gauge or mass terms, the surface Dirac Hamiltonian in the local orthonormal frame takes the standard form
\begin{equation}
    H
    = v_{\rm F}\,\bm{\sigma}\cdot\bm{p}
    = v_{\rm F}\bigl(\sigma_1 p_1 + \sigma_2 p_2\bigr),
\end{equation}
where $(\sigma_1,\sigma_2)$ are Pauli matrices in the $\{e^1,e^2\}$ basis. Using Eq.~\eqref{eq:local_momenta}, the energy spectrum becomes
\begin{equation}
    E_{m}(k_z)
    = \pm v_{\rm F}\sqrt{p_1^2 + p_2^2}
    = \pm \hbar v_{\rm F}
      \sqrt{\left( \frac{m}{R} - \Omega R k_z \right)^2 + k_z^2}.
    \label{eq:Dirac_dispersion_TI}
\end{equation}
This expression is the electronic analogue of the geometric circular birefringence derived for photons: the uniform torsion parametrized by $\Omega$ induces a momentum-space mixing between the azimuthal quantum number $m$ and the longitudinal wave vector $k_z$.

\subsection{Torsion-induced momentum mixing and relation to dislocation-line modes}

It is instructive to rewrite Eq.~\eqref{eq:Dirac_dispersion_TI} in terms of a dimensionless torsion parameter
\begin{equation}
    \tau = \Omega R^2,
    \qquad
    \kappa = k_z R,
\end{equation}
so that
\begin{equation}
    E_{m}(\kappa)
    = \pm \frac{\hbar v_{\rm F}}{R}
      \sqrt{\bigl(m - \tau\kappa\bigr)^2 + \kappa^2}.
    \label{eq:Dirac_dispersion_dimensionless}
\end{equation}
In the limit $\tau \to 0$ we recover the familiar dispersion of Dirac surface states on a flat cylinder,
\begin{equation}
    E_{m}^{(0)}(\kappa)
    = \pm \frac{\hbar v_{\rm F}}{R}
      \sqrt{m^2 + \kappa^2},
\end{equation}
with decoupled quantization along $\phi$ and continuous motion along $z$. For finite torsion, the combination $(m,\kappa)$ is sheared into $(m - \tau\kappa,\kappa)$, indicating that the effective azimuthal momentum experienced by the Dirac fermions depends linearly on $k_z$. This momentum-space shear can be viewed as a geometric analogue of the Aharonov--Bohm phase associated with dislocations in TIs and Chern insulators \cite{TeoKane2010_PRB,Juricic2012_PRL,Slager2014_PRB}: the torsion plays the role of an emergent background field that couples to the Dirac cone.

From a topological perspective, dislocation lines in strong TIs are known to host one-dimensional modes whenever the Burgers vector has a nontrivial projection onto the weak topological indices \cite{TeoKane2010_PRB,RanVishwanath2009_NatPhys,Slager2014_PRB}. The dispersion~\eqref{eq:Dirac_dispersion_dimensionless} provides a complementary continuum description in which a macroscopic density of screw dislocations, encoded in $\tau$, continuously deforms the surface Dirac cone rather than producing isolated line modes. In the limit of large $|\tau|$ and for fixed $m$, Eq.~\eqref{eq:Dirac_dispersion_dimensionless} can be reorganized as
\begin{equation}
    E_{m}(\kappa)
    \simeq \pm \frac{\hbar v_{\rm F}|\tau|}{R}
           \sqrt{\left( \kappa - \frac{m}{\tau} \right)^2
                 + \mathcal{O}\!\left(\frac{1}{\tau^2}\right)},
\end{equation}
showing that the Dirac cone is effectively tilted and shifted in $(m,\kappa)$ space by an amount controlled by the torsion. This behaviour is reminiscent of the emergent pseudo-gauge fields and torsional responses discussed in Weyl semimetals with dislocations \cite{SumiyoshiFujimoto2016_PRL}, and suggests that torsion-engineered TI surfaces could provide a solid-state realization of elastic gauge fields closely related to the geometric optical activity analysed in this work.

A full treatment of the electronic problem would include spin connections, possible mass terms, and coupling to external electromagnetic fields, and could reveal additional phenomena such as torsional Landau quantization or torsion-induced chiral currents. Nevertheless, the simple spectrum~\eqref{eq:Dirac_dispersion_TI} already captures the essential message: the spiral geometry that produces circular birefringence for light also imprints a universal, geometry-induced anisotropy on Dirac surface fermions, opening a direct route to connect the present optical model with the rich physics of dislocations and topological responses in condensed-matter systems.
\begin{figure*}[tbhp]
\centering
\includegraphics[scale=0.5]{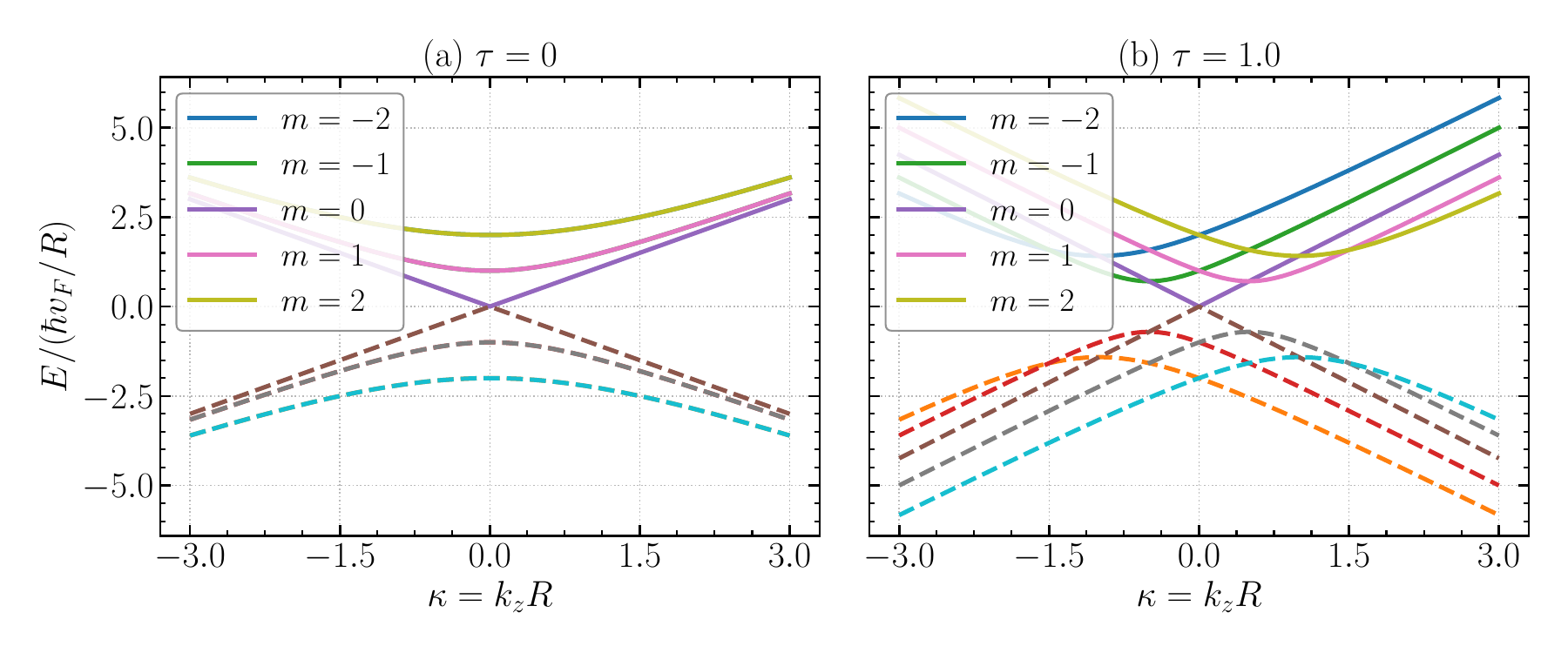}
\caption{Dimensionless dispersion relation for Dirac surface states
on a cylindrical topological insulator in the presence of torsion.The longitudinal momentum and energy are rescaled as $\kappa = k_z R$ and $E / (\hbar v_F / R)$, respectively.
Panel~(a) shows the torsionless case ($\tau = \Omega R^2 = 0$), where
the branches $E_m^{(0)}(\kappa) = \pm \sqrt{m^2 + \kappa^2}$ form
a symmetric Dirac cone. Panel~(b) displays the spectrum for a finite
torsion parameter ($\tau = 1$), where the dispersion becomes $E_m(\kappa) = \pm \sqrt{(m - \tau\kappa)^2 + \kappa^2}$ and the
cone is sheared in momentum space. Different colors correspond to azimuthal quantum numbers $m = 0, \pm 1, \pm 2$. The torsion-induced deformation of the cone is the electronic analogue of the geometric birefringence obtained for photons in the spiral geometry.}
\label{fig:TI_spectrum}
\end{figure*}
To make the electronic analogue of the geometric optical activity more explicit, we now examine the spectrum of Dirac surface states on a cylindrical topological insulator in the presence of torsion. Using the effective surface Hamiltonian in the spiral geometry, we obtain the dimensionless dispersion relation
\begin{equation}
    \frac{E_m(\kappa)}{\hbar v_F/R}
    = \pm \sqrt{(m - \tau \kappa)^2 + \kappa^2},
    \label{eq:TI_dispersion}
\end{equation}
where $\kappa = k_z R$, $m \in \mathbb{Z}$ is the azimuthal quantum
number, $v_F$ is the Fermi velocity, and $\tau = \Omega R^2$ is the
dimensionless torsion parameter inherited from the screw-dislocation
density. In the absence of torsion ($\tau = 0$), Eq.~\eqref{eq:TI_dispersion}
reduces to the familiar cylindrical Dirac cone
$E_m^{(0)}(\kappa) = \pm \sqrt{m^2 + \kappa^2}$.

These features are illustrated in Fig.~\ref{fig:TI_spectrum}. In
panel~\ref{fig:TI_spectrum}(a), the torsionless spectrum exhibits a
symmetric cone centered at $\kappa = 0$, with equally spaced subbands
labelled by $m$. Panel~\ref{fig:TI_spectrum}(b) shows the effect of a
finite torsion parameter ($\tau = 1$): the branches $E_m(\kappa)$ become
asymmetric, reflecting the replacement $m \to m - \tau\kappa$ in
Eq.~\eqref{eq:TI_dispersion}. From the geometric point of view, torsion
“shears’’ the Dirac cone along the $\kappa$ direction, in close analogy
with the splitting of the circularly polarized photonic modes induced by
the same torsion parameter in Sec.~\ref{sec:design_rules}. This spectral
deformation provides a direct electronic counterpart of the torsion-induced
circular birefringence discussed for light: in both cases, uniform
torsion lifts a degeneracy by coupling an internal degree of freedom
(polarization or angular momentum) to the longitudinal propagation
direction.
\begin{figure*}[tbhp]
    \centering
    \includegraphics[scale=0.5]{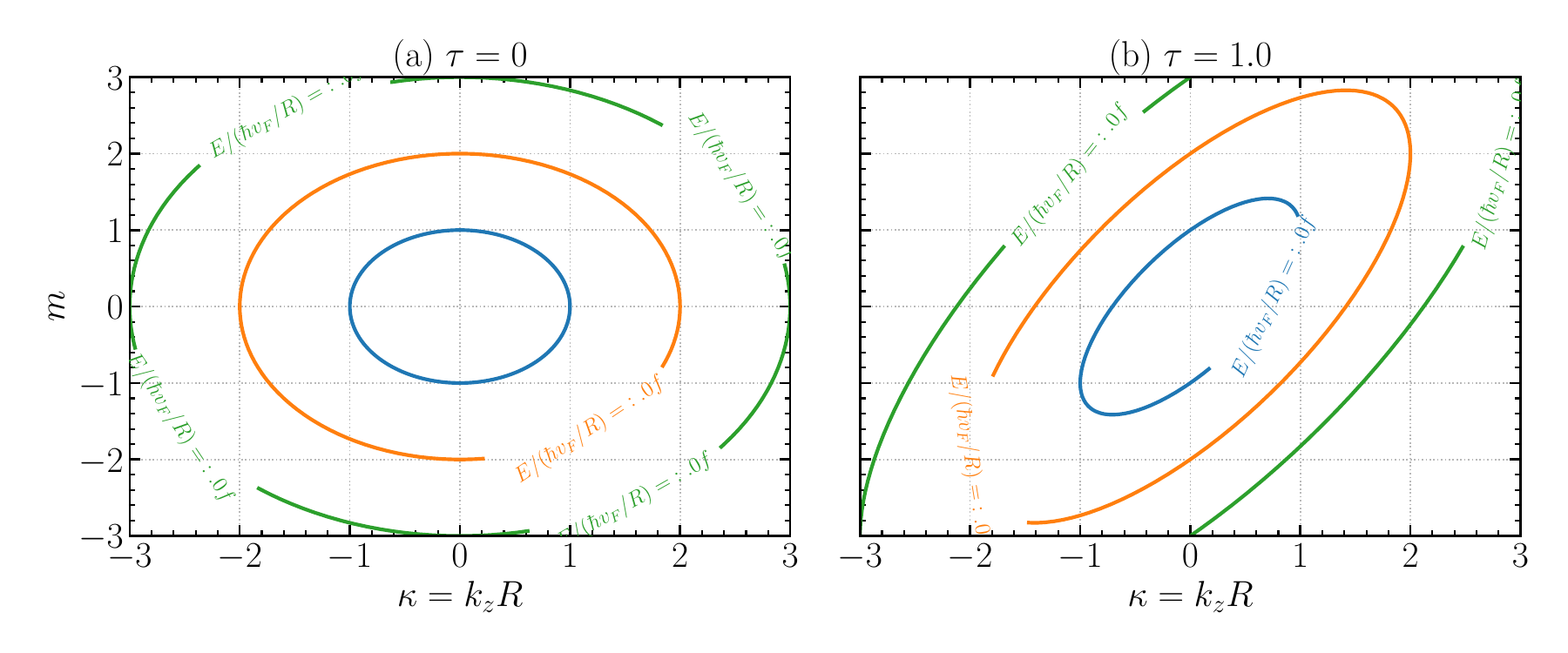}
    \caption{Constant-energy contours for Dirac surface states on a
    cylindrical topological insulator in the $(\kappa,m)$ plane, with
    $\kappa = k_z R$ and $E / (\hbar v_F / R)$ fixed.
    Panel~(a) shows the torsionless case ($\tau = 0$), where the contours
    $E^2 = m^2 + \kappa^2$ are circular and reflect the isotropic Dirac
    cone. Panel~(b) shows the case with finite torsion ($\tau = 1$), where
    the dispersion $E^2 = (m - \tau\kappa)^2 + \kappa^2$ leads to sheared
    contours in momentum space. The torsion parameter thus acts as an effective
    elastic gauge field that mixes longitudinal and azimuthal motion,
    in close analogy with emergent gauge fields generated by dislocations
    in topological semimetals and topological insulators.}
    \label{fig:TI_contours}
\end{figure*}
Further insight into the role of torsion is provided by inspecting
constant-energy cuts of Eq.~\eqref{eq:TI_dispersion} in the
$(\kappa,m)$ plane. For fixed dimensionless energy
$E/(\hbar v_F/R)$, the torsionless dispersion
$E^2 = m^2 + \kappa^2$ describes circular contours, emphasizing
the isotropic nature of the Dirac cone in the $(\kappa,m)$ variables.
In contrast, a finite torsion parameter $\tau$ deforms these circles into
sheared curves defined by
$E^2 = (m - \tau\kappa)^2 + \kappa^2$.

This behavior is illustrated in Fig.~\ref{fig:TI_contours}. In
panel~\ref{fig:TI_contours}(a), the contours at selected energies are
almost perfectly circular, confirming the isotropic coupling between
longitudinal and azimuthal motion when $\tau = 0$. In
panel~\ref{fig:TI_contours}(b), the same energy levels become tilted ellipses
once $\tau$ is turned on, revealing a linear mixing between $m$ and $\kappa$.
From the perspective of effective field theory, the torsion parameter plays
the role of an elastic gauge field that shifts $m \to m - \tau\kappa$ and
thereby distorting the Fermi surface in momentum space. This is the same
mechanism by which dislocations and strain generate emergent gauge fields in
topological semimetals and topological insulators, now reinterpreted as the
electronic counterpart of the torsion-induced geometric birefringence derived
for photons in Sec.~\ref{sec:design_rules}.
\begin{table*}[t]
    \centering
    \caption{Geometric dictionary between the torsion-induced optical
    activity for photons in the spiral geometry and the torsion-induced
    deformation of the Dirac surface states in a cylindrical topological
    insulator. The same torsion parameter $\Omega$, generated by a
    continuous density of screw dislocations, controls both the
    circular birefringence of light and the shear of the Dirac cone
    in momentum space.}
    \label{tab:geom_dictionary}
    \vspace{0.2cm}
    \begin{tabular}{lll}
        \toprule
        \textbf{Quantity / concept} &
        \textbf{Optical model (this work)} &
        \textbf{Topological-insulator surface analogue} \\
        \midrule
        Background field &
        Uniform torsion $\Omega$ in spiral geometry &
        Uniform torsion $\Omega$ from screw-dislocation density \\[2pt]
        Longitudinal coordinate &
        $z$ (beam propagation direction) &
        $z$ (cylindrical surface direction) \\[2pt]
        Transverse scale &
        Beam radius $\rho$ &
        Cylinder radius $R$ \\[2pt]
        Wave number/momentum &
        $k_z^{(\pm)}$ (RCP/LCP modes) &
        $k_z$ (longitudinal momentum) \\[2pt]
        Angular index &
        Circular polarization helicity ($\pm$) &
        Azimuthal quantum number $m$ \\[2pt]
        Torsion parameter (dimensionless) &
        $\Omega\rho L$ (rotation phase) &
        $\tau = \Omega R^2$ (shear strength in $E_m$) \\[2pt]
        Splitting/mixing &
        $\Delta k_z = 2\Omega\rho$ &
        Momentum mixing $(m - \tau \kappa)$, $\kappa = k_z R$ \\[2pt]
        Main observable &
        Polarization rotation
        $\Delta\theta = \Omega\rho L$ &
        Dirac-cone deformation
        $E_m(\kappa) = \pm \sqrt{(m - \tau\kappa)^2 + \kappa^2}$ \\
        \bottomrule
    \end{tabular}
\end{table*}
The geometric correspondence summarized in Table~\ref{tab:geom_dictionary}
highlights how the same torsion parameter $\Omega$, generated by a continuous distribution of screw dislocations, controls both the optical and electronic responses of the system. On the optical side, the spiral Riemann--Cartan geometry provides a uniform torsion background in which light propagates; on the electronic side, the same type of torsion emerges effectively on the surface of a cylindrical topological insulator or in a dislocation-rich
Dirac/Weyl material. In both cases, torsion is not an external field in the usual electromagnetic sense, but a geometric property encoded in the underlying defect distribution.

The second and third rows emphasize the common role of the longitudinal coordinate $z$ and of a characteristic transverse scale (the beam radius $\rho$ for photons and the cylinder radius $R$ for Dirac surface states). In the optical problem, $z$ sets the propagation direction along which the polarization rotation accumulates, whereas $\rho$ selects the transverse position where the torsion-induced birefringence is evaluated. In the
topological-insulator analogue, $z$ plays the role of the propagation coordinate on the cylindrical surface, while $R$ fixes the scale of the azimuthal quantization and enters the dimensionless variables $\kappa = k_z R$ and $\tau = \Omega R^2$.

The fourth and fifth rows connect the relevant “internal” and kinematic degrees of freedom in each system. For photons, the quantities $k_z^{(\pm)}$ label the longitudinal wavenumbers of right- and left-circularly polarized modes, and the helicity (RCP/LCP) is the internal index that couples to torsion. For Dirac surface states, $k_z$ is the longitudinal momentum and $m$ is the azimuthal quantum number, which plays a role analogous to the polarization index: it is this discrete angular-momentum label that becomes mixed with the longitudinal motion through the torsion-induced shift $m \to m - \tau \kappa$.

The next two rows show explicitly how the torsion strength enters in dimensionless form. In the optical case, the central combination is $\Omega \rho L$, which directly controls the geometric phase accumulated by a polarization component after traversing a distance $L$ at radius $\rho$. This same quantity sets the polarization-rotation angle $\Delta\theta = \Omega \rho L$. In the electronic problem, the torsion parameter appears as $\tau = \Omega R^2$, which measures the effective shear of the Dirac cone in the $(\kappa,m)$ plane. Increasing $\tau$ enhances the mixing between longitudinal and azimuthal motion and thus the deformation of
the spectrum.

Finally, the last row makes the analogy between observables explicit. On the photonic side, the key effect is a torsion-induced splitting $\Delta k_z = 2 \Omega \rho$ between the two circular polarizations, which leads to the geometric optical activity described in Sec.~\ref{sec:design_rules} through the rotation angle $\Delta\theta = \Omega \rho L$. On the electronic side, the same torsion parameter produces a shear of the Dirac cone, encoded
in the dispersion
\begin{equation}
    E_m(\kappa) = \pm \sqrt{(m - \tau \kappa)^2 + \kappa^2},
\end{equation}
which deforms constant-energy contours and effectively acts as an elastic gauge field in momentum space. The table thus makes transparent that circular birefringence for photons and Dirac-cone distortion for electrons are two manifestations of the same geometric mechanism: the coupling between an internal degree of freedom (polarization or angular momentum) and a uniform torsion background generated by screw dislocations.

\section{Conclusions}

We have performed an explicit derivation of electromagnetic wave propagation in a spacetime with uniform torsion, starting from the covariant Maxwell equations in a Riemann-Cartan geometry. By identifying the torsion-dependent terms in the wave operator, we showed that a single contortion component, proportional to the screw-dislocation density, produces a chiral coupling between the polarization components of the electromagnetic field. The resulting dispersion relation lifts the degeneracy between right- and left-circularly polarized modes, giving rise to a purely geometric circular birefringence. In the paraxial, small-torsion regime, this effect leads to a remarkably simple law for the rotation of the polarization plane of a linearly polarized beam,
\begin{equation}
    \Delta\theta = \Omega\rho L,
\end{equation}
which is linear in the defect density $\Omega$, accumulates with the propagation distance $L$, and depends on the transverse coordinate $\rho$. This radius-dependent rotation is a distinctive prediction of the spiral geometry and provides a direct optical signature of uniform torsion.

Recasting our results in the standard language of optical activity, we identified an effective geometric birefringence
$\Delta n = 2c\Omega\rho/\omega$ and a corresponding geometric rotatory power that are fully determined by the Burgers-vector density and the beam position. Using structural parameters extracted from the literature for dislocated semiconductors such as GaN, ZnO, SiC, and related materials, we obtained birefringence values in the range $10^{-9}$–$10^{-8}$ and polarization rotations of a few millidegrees over sub-millimetre and millimetre path lengths. Although numerically small compared with strongly anisotropic crystals, these effects lie within the sensitivity of modern high-precision polarimetric setups and can be significantly enhanced in engineered dislocation networks, structured waveguides, or metamaterials whose unit cells emulate the spiral geometry.

From a quantum-optical perspective, the torsion-induced splitting between circular polarizations defines a broadband geometric phase gate for polarization qubits, described by the unitary
$U(\Omega,\rho,L) = \exp[-i\,\Omega\rho L\,\sigma_3]$. We analysed the gate fidelity under realistic fluctuations of the geometric parameters and showed that infidelities below $10^{-3}$ are compatible with percent-level control of the effective torsion and sample thickness, indicating that torsion-mimicking structures can act as robust, passive $Z$ gates in integrated photonics and chiral quantum-optical architectures.

Finally, we constructed an electronic analogue of the same geometric mechanism by considering Dirac surface states of a cylindrical topological insulator embedded in the spiral geometry. In this setting, the torsion parameter $\Omega$ mixes longitudinal and azimuthal momenta and shears the surface Dirac cone according to
\begin{equation}
    E_m(\kappa) = \pm \frac{\hbar v_F}{R}\sqrt{(m - \tau\kappa)^2 + \kappa^2},
\end{equation}
with $\kappa = k_z R$ and $\tau = \Omega R^2$. The deformation of the Dirac cone and the associated distortion of constant-energy contours are the electronic counterpart of the circular birefringence derived for photons, and can be interpreted as an elastic gauge-field response to screw dislocations. In this way, the spiral geometry provides a unified geometric framework in which torsion-induced optical activity, polarization-phase control, and topological electronic responses are different manifestations of the same underlying torsion field. This connection suggests concrete routes for probing torsion in dislocation-rich semiconductors, topological insulators, and metamaterials, and for exploiting geometric design principles in both classical and quantum photonic devices.

\section*{Acknowledgments}

This work was supported by CAPES (Finance Code 001), CNPq (Grant 306308/2022-3), and FAPEMA (Grants UNIVERSAL-06395/22 and APP-12256/22).

\bibliographystyle{apsrev4-2}
%


\end{document}